\documentclass[aps,twocolumn,superscriptaddress]{revtex4}

\usepackage{amsmath,amssymb,bm}
\usepackage{graphicx}
\usepackage{amsfonts}

\newcommand{\be}{\begin{equation}}
\newcommand{\ee}{\end{equation}}

\newcommand{\ba}{\begin{eqnarray}}
\newcommand{\ea}{\end{eqnarray}}

\begin{document}

%%%%%%%%%%%%%%%%%%%%%%%%%%%%%%%%%%%%%%%%%%%
%%%%%%%%%%%%%%%% the text %%%%%%%%%%%%%%%%%
%%%%%%%%%%%%%%%%%%%%%%%%%%%%%%%%%%%%%%%%%%%
\title{Gluonic Structure of the Constituent Quark}

\author{Nikolai Kochelev}
%\email{kochelev@theor.jinr.ru}
 \affiliation{Institute of Modern
Physics, Chinese Academy of Science, Lanzhou 730000, China}
\affiliation{Bogoliubov Laboratory of Theoretical Physics, Joint
Institute for Nuclear Research,\\ Dubna, Moscow Region, 141980
Russia}

\author{Hee-Jung Lee}
% \email{hjl@chungbuk.ac.kr}
\affiliation{Department of Physics Education, Chungbuk National
University, Cheongju, Chungbuk 361-763, Korea}

\author{Baiyang Zhang}
%\email{zhangbaiyang@impcas.ac.cn}
 \affiliation{Institute of Modern
Physics, Chinese Academy of Science, Lanzhou 730000, China}
\author{Pengming Zhang}
%\email{zhpm@impcas.ac.cn}
 \affiliation{Institute of Modern Physics,
Chinese Academy of Science, Lanzhou 730000, China}

\begin{abstract} Based on both the constituent quark picture and the instanton model for QCD vacuum,
we calculate the unpolarized and polarized gluon distributions in the constituent quark and in the nucleon.
Our approach consists of the two main steps. At the first step, we calculate the
gluon distributions inside the constituent quark generated by the perturbative quark-gluon
interaction, the non-perturbative quark-gluon interaction, and the non-perturbative quark-gluon-pion anomalous
chromomagnetic interaction. The non-perturbative interactions are related to
the existence of the instantons, strong topological fluctuations
of gluon fields, in the QCD vacuum. At the second step, the convolution model is applied to
derive the gluon distributions in the nucleon.
A very important role of the pion field in producing the
unpolarized and the polarized gluon distributions in the hadrons is discovered.
We discuss a possible solution of the proton spin problem.

\end{abstract}
\maketitle

\section{Introduction}
The parton distribution functions (PDFs) are one of the cornerstones of the calculation
of high energy cross sections. Among the various PDFs, it is very important to know the
gluon distribution function in the proton for understanding especially the experiments
at Large Hadron Collider (LHC) where the gluons give the dominant contribution to the different cross
sections. Furthermore, the careful calculation of the polarized gluon distribution
in the proton within an approach based on the modern model for the non-perturbative effects in QCD
might give a solution of the the so-called proton spin crisis problem (for the recent review see \cite{Aidala:2012mv}).
\\
In this Letter we calculate the unpolarized and polarized gluon
distributions in the constituent quark and in the nucleon.
Our calculation is based on both the constituent quark model for the nucleon and
the instanton model for the non-perturbative QCD vacuum (see reviews\cite{shuryak,diakonov}).
The instantons, the strong nonperturbative fluctuations of the vacuum gluon fields, describe
non-trivial topological structure of the QCD vacuum and they give a natural explanation
of the fundamental phenomena in the strong interaction such as the spontaneous chiral symmetry breaking (SCSB)
and $U(1)_A$ symmetry violation. The average size of the instantons $\rho_c\approx 1/3$~fm is
much smaller than the confinement size $R_c\approx 1$~fm and can be considered as the scale of SCSB.
Furthermore, SCSB induced by the instantons is responsible for the formation of the constituent
massive quark with the size $R_q\approx \rho_c\approx 1/3$~fm. The SCSB plays very important role
not only in hadron spectroscopy but also in the reactions with hadrons. In particulary,
the specific mechanism based on SCSB is needed for the explanation of a large quark spin-flip effects
observed in high energy reactions (see, for example, the discussion in \cite{kochelev3}).
The mechanism for this phenomenon coming from an anomalous chromomagnetic quark-gluon
interaction induced by the instantons was suggested in \cite{kochelev1} by one of the authors
of this Letter. The anomalous quark chromomagnetic moment (AQCM) $\mu_a$, which
is proportional to the parameter of the SCSB violation $\delta=(M_q\rho_c)^2$ where $M_q$ is
the so-called effective mass of the quark in instanton vacuum \cite{diakonov,kochelev4},
contributes to the high energy reaction so that can be responsible for the observed large spin effects.
To preserve the partial conservation of the axial-vector current (PCAC), the modification of this interaction
by including the pion field into consideration was proposed in \cite{Balla:1997hf,diakonov}.
Recently it was demonstrated that the anomalous chromomagnetic quark-gluon-pion interaction
gives a very important contribution to the high energy inclusive pion production
in the proton-proton collisions \cite{Kochelev:2015pha}. This result could be a solution of the
longstanding problem in the description of the pion inclusive cross section in the fixed target experiments.
Moreover it was shown that this interaction might be responsible for the large energy loss of the fast partons
in quark-gluon plasma \cite{Kochelev:2015jba}.
\\
The main goal of our Letter is to calculate the contributions from the anomalous quark-gluon and quark-gluon-pion
interactions to the unpolarized and polarized gluon distributions inside the constituent quark and in the nucleon.
We will also discuss the contribution coming from the perturbative quark-gluon interaction to these gluon distributions.

\section{Gluon unpolarized and polarized distributions in the constituent quark}

The effective Lagrangian based on the quark-gluon chromomagnetic interaction which preserves the  chiral symmetry
has the following form \cite{Balla:1997hf,diakonov}
\begin{equation}
{\cal L}_I= -i\frac{g_s\mu_a}{4M_q}\bar q\sigma^{\mu\nu}t^a
e^{i\gamma_5\vec{\tau}\cdot\vec{\phi}_\pi/F_\pi}q G^{a}_{\mu\nu},
\label{Lag2}
\end{equation}
where $\mu_a$ is AQCM\footnote{The present definition of AQCM is corresponded to the relation
$\mu_a=(g-2)/2$, where $g$ is Lande's factor. It is different by the factor 2 from what was used in
the papers \cite{kochelev1} and \cite{diakonov}.}, $g_s$ is the strong
coupling constant, $G^{a}_{\mu\nu}$ is the gluon field strength, and $F_\pi = 93~\mbox{MeV}$ is the pion decay constant.
 In the first order of the pion field, the  Lagrangian reads as \cite{Kochelev:2015pha,Kochelev:2015jba}
\begin{equation}
\mathcal{L}_I =- i\frac{g_s\mu_a}{4M_q} \, \bar q\sigma^{\mu\nu}t^a q \, G^{a}_{\mu\nu}
+\frac{g_s\mu_a}{4M_qF_\pi}\bar q \sigma^{\mu\nu} t^a \gamma_5
\bm{\tau}\cdot \bm{\pi} q \, G^{a}_{\mu\nu}.
\label{Lag}
\end{equation}
 The multi-instanton effects are presented in this Lagrangian by the
effective quark mass $M_q$  in the quark zero-mode-like propagator in the instanton field \cite{Faccioli:2001ug}.
In principle, the additional multi-instanton effects might be possible but the strong suppression
of them is expected due to the small
packing fraction of the instantons in the QCD vacuum $f\approx 1.5\times 10^{-2}$ \cite{shuryak}.
Within the instanton model, the value of AQCM is \cite{diakonov,kochelev4}
\be
\mu_a=-\frac{3\pi (M_q\rho_c)^2}{4\alpha_s(\rho_c)}.
\label{AQCM1}
\ee
We will fix the value of the strong coupling constant at instanton scale as
$\alpha_s(\rho_c)=g_s^2(\rho_c)/4\pi \approx 0.5$ \cite{diakonov}.
Using Eq.~\ref{AQCM1}, the second term of Eq.~\ref{Lag}, which presents the quark-gluon-pion interaction, can be
rewritten as \cite{Kochelev:2015jba}
\begin{equation}
{\cal L}_{\pi qqg}=-\frac{3\pi^2\rho_c^2}{4g_s(\rho_c)}g_{\pi qq}\,
\bar q\sigma^{\mu\nu}t^a \gamma_5 \bm{\tau}\cdot \bm{\pi} q \, G^{a}_{\mu\nu},
\label{vertex}
\end{equation}
where $g_{\pi qq}=M_q/F_\pi$ is the quark-pion coupling constant.

We will follow  the Altarelli-Parisi (AP) approach \cite{Altarelli:1977zs} to calculate
the gluon distribution functions in the constituent quark. The contributions from
the perturbative QCD (pQCD) and the non-perturbative instanton-induced interactions are shown in Fig.1.
\begin{figure}[htbp]
\centering{\includegraphics[width=9cm,height=2.5cm,angle=0]{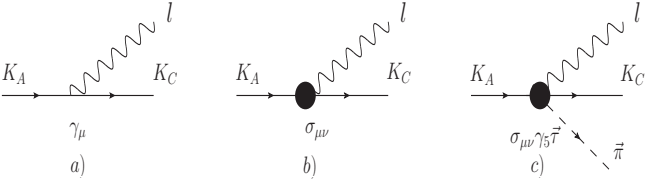}}
\caption{a) corresponds to the contributions from the pQCD to gluon distribution in the quark.
b) and c) correspond to the contributions to the gluon distribution in the quark
from the non-perturbative quark-gluon interaction and the non-perturbative quark-gluon-pion interaction, respectively.}
\label{fig:1}
\end{figure}
In the single instanton approximation the expansion parameter is $\delta = (M_q\rho_c)^2 $.
For the value of the quark mass $M_q=86 $ MeV in~\cite{Faccioli:2001ug},
this parameter is small $\delta \approx 0.02 $. Therefore, we can neglect quark mass
in the infinite momentum frame $P\rightarrow\infty $. In this frame, the momenta of the partons and
the polarization vectors of the gluon for the diagrams a) and b) in Fig.~1 are
\ba
K_A&=&(P,P,0),
\nonumber\\
K_C&=&((1-z)P+\frac{\vec{p}_\perp^2}{2P(1-z)},(1-z)P,-\vec{p}_\perp),
\nonumber\\
& l & =(zP-\frac{\vec{p}_\perp^2}{2P(1-z)},zP,\vec{p}_\perp),
\nonumber\\
\epsilon_{\pm} &=&\frac{1}{\sqrt{2}}(0,-\frac{p_x\pm ip_y}{zP},1,\pm i).
\label{momenta1}
\ea
The contribution from pQCD to the unintegrated gluon non-polarized and polarized distributions
in the quark are given by~\cite{Altarelli:1977zs}
\ba
dg(z,p_\perp^2)&=&
C_F\frac{\alpha_s(\mu^2)}{2\pi}\frac{1+(1-z)^2}{z}\frac{dp_\perp^2}{p_\perp^2},
\nonumber\\
d\Delta g^{pert} (z,p_\perp^2)&=&
C_F\frac{\alpha_s(\mu^2)}{2\pi}(2-z)\frac{dp_\perp^2}{p_\perp^2},
\label{gluonpQCD}
\ea
where $C_F=4/3$ is the color factor.
The non-perturbative vertex b) in Fig.1\footnote{The first attempt to calculate the contribution
to the gluon distribution coming from the diagram Fig.1b was done in \cite{Kochelev:1997qq}. Unfortunately,
the incorrect kinematics was used in that paper.}, in which the helicity of the quark is flipped,
produces the following unpolarized unintegrated gluon distribution
 \be
 dg^{nonpert,b} (z,p_\perp^2)=
 C_F\frac{\alpha_s(\mu^2)}{2\pi}\bigg(\frac{\mu_a}{2M_q}\bigg)^2 \frac{2}{z}F_g^2(|l|\rho_c)dp_\perp^2,
 \label{gluonnonpQCD}
 \ee
where $F_g(x)=4/x^2-2K_2(x)$ is the instanton form factor \cite{kochelev1,kochelev4} and
$|l|=|p_\perp|/\sqrt{1-z}$.
One can see that due to the flip of the quark's helicity the factor $p_\perp^2$ is absent
in the denominator of Eq.\ref{gluonnonpQCD} in comparison with pQCD case Eq.\ref{gluonpQCD}.
Therefore, the non-perturbative contribution gives the rise to the distribution of the gluon with the large $ p_\perp^2$.
Using Eq.\ref{AQCM1}, Eq.\ref{gluonnonpQCD} can be rewtitten in more compact form
\be
dg^{nonpert,b} (z,p_\perp^2)=
|\mu_a| \frac{1}{4z}F_g^2(|l|\rho_c)dp_\perp^2,
\label{gluonnonpQCD1}
\ee
where the natural scale $\mu^2=1/\rho_c^2$ was chosen in the running strong coupling constant.
For the integrated gluon distribution we have
\be
g^{nonpert,b} (z,Q^2)=
|\mu_a| \frac{1-z}{2z}\int_{s_{min}}^{s_{max}} dssF_g^2(s),
\label{gluonnonpQCD1}
\ee
with upper limit $s_{max}=|Q|\rho_c/(1-z)$, where $Q^2$ is the external virtuality.
 For the low limit we should use $s_{min}=|{p_\perp}_{min}|\rho_c/(1-z)$, where
$|{p_\perp}_{min}|$ is fixed by the confinement scale $|{p_\perp}_{min}|\approx \Lambda_{QCD}\approx 200 $ MeV.
However in our calculation below for $Q^2\gg   \Lambda_{QCD}^2$ we can safety put $|{p_\perp}_{min}|\rightarrow 0$.
In this case,  for the limit $Q^2\rightarrow \infty $, our result  
\be
g^{nonpertb}(z)=
|\mu_a| \frac{1-z}{z}
\label{gluonnonpQCD1}
\ee
is in agreement with \cite{diakonov}
The contribution of the diagram b) in Fig.1 to the polarized gluon distribution
is vanished. This result of gluon polarization coming from quark-gluon chomomagnetic vertex without pion
is also in agreement with statement presented in \cite{diakonov}.
\\
The non-perturbative contribution with the pion, the diagram c) in Fig.1, has never been
considered. We will show that this contribution determines the behavior
of the unpolarized and polarized gluon distributions in the constituent quark at small $z$
and in the nucleon at small Bjorken's $x$. For this diagram, the momenta of the partons are
\ba
K_A&=&(P,P,0),
\nonumber\\
K_C&=&((1-x-z)P+\frac{(\vec{p}_\perp+\vec{k}_\perp)^2}{2P(1-x-z)},
\nonumber\\
&&(1-x-z)P,-(\vec{p}_\perp+\vec{k}_\perp)),
\nonumber\\
l &=&(zP-\bigg[\frac{\vec{k}_\perp^2}{2xP}+\frac{(\vec{p}_\perp+\vec{k}_\perp)^2}{2P(1-x-z)}\bigg],zP, \vec{p}_\perp),
\nonumber\\
K_\pi&=&(xP+\frac{\vec{k}_\perp^2}{2xP},xP, \vec{k}_\perp).
\label{momenta2}
\ea
The small mass of pion is neglected in Eq.\ref{momenta2}.
Since the three particles are appeared in the final state, the integration over the additional phase space
should be done as
\be
\bigg[\frac{d^3k_C}{(2\pi)^32E_C}\bigg]^{without\  pion}\rightarrow
\bigg[\frac{d^3k_C}{(2\pi)^32E_C}\frac{d^3k_\pi}{(2\pi)^32E_{k_\pi}}\bigg]^{with\  pion}
\nonumber
\ee
By using Eq.\ref{momenta2} this phase space factor can be transformed into
\be
dPS=\frac{1}{2^8\pi^6}\frac{dxdz}{x(1-x-z)}d\vec{k}_\perp d\vec{p}_\perp.
\label{PS}
\ee
Calculating the matrix element for the diagram c) in Fig.1 and performing of the integration over $x$,
the unintegrated unpolarized and polarized gluon distribution in the quark are
\ba
&& dg^{nonpert,c}(z,p_\perp^2)=C\int_{y_{min}}^{y_{max}}dy F(t,y,z)\nonumber\\
&&\times
\frac{F_g^2(\sqrt{zy+t})}{(zy+t)^2} dp_\perp^2,\nonumber\\
&&d\Delta g^{nonpert,c}(z,p_\perp^2)=C\int_{y_{min}}^{y_{max}}dy \Delta F(t,y,z)
\nonumber\\
&&\times
\frac{F_g^2(\sqrt{zy+t})}{(zy+t)^2} dp_\perp^2,
\label{withpion1}
\ea
where
\be
C=\frac{3C_F}{\alpha_s(\rho_c)}\frac{9}{2^{12}\pi}g^2_{\pi qq}\rho_c^2.
\label{constant}
\ee
The first factor 3 in Eq.\ref{constant} is related to the isospin.
The $F(t,y,z)$ and $\Delta F(t,y,z) $ functions in Eq.\ref{withpion1} are given by
\ba
&&F(t,y,z)=\frac{1-z}{z}(2t^2+y^2z^2+2zty+z^2ty) \nonumber\\
&&\Delta F(t,y,z)=(1-z)y(2t-zt+zy),
\label{func}
\ea
where $t=p_\perp^2\rho_c^2/(1-z)$, $y=M_X^2\rho_c^2/(1-z)$ and
$M_X^2=(k_C+k_\pi)^2$ is the invariant mass of the final pion-quark system.
As the low limit for the integration over $y$ the value  $M_X^{min}=M_q+m_\pi$
with $m_\pi=140$ MeV is used and as the upper limit
$M_X^{max}=E_{sph}$, where
$E_{sph}=3\pi/(4\alpha_s(\rho_c)\rho_c)$ is so-called sphaleron energy
(see discussion in \cite{Kochelev:2015pha}), is used.
The integrated  gluon distributions in the quark are defined by
\ba
g(z,Q^2)&=&\int_{{p_\perp^2}_{min}}^{{p_\perp^2}_{max}} dg(z,p_\perp^2),
\nonumber\\
\Delta g(z,Q^2)&=&\int_{{p_\perp^2}_{min}}^{{p_\perp^2}_{max}} d\Delta g(z,p_\perp^2).
\label{int}
\ea
The integration limits in Eq.\ref{int} for the non-perturbative case are ${p^2_\perp}_{min} =0$
and ${p^2_\perp}_{max}=Q^2$. For the perturbative case, it is more natural to use the low limit of integration
${p^2_\perp}_{min}=1/\rho_c^2$ because for  $p^2_\perp \leq 1/\rho_c^2$ one cannot believe in the
validity of pQCD. That means that in our model the perturbative unpolarized and polarized
gluon distributions in the constituent quark vanish if the external momentum
$Q^2\leq 1/\rho_c^2=0.35$ GeV$^2$.

\begin{widetext}

\begin{figure}[h]
\begin{minipage}[c]{6cm}
%\vskip -0.5cm
\hspace*{-3.0cm}
\centering{\includegraphics[width=5cm,height=3cm,angle=0]{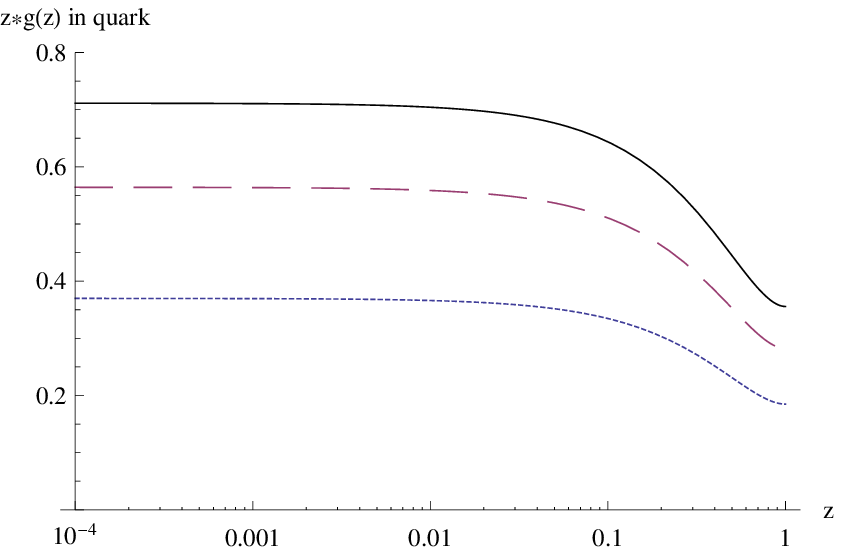}}\
\end{minipage}
\begin{minipage}[c]{8cm}
\hspace*{-5.0cm}
 %\vskip -1cm
\centering{\includegraphics[width=5cm,height=3cm,angle=0]{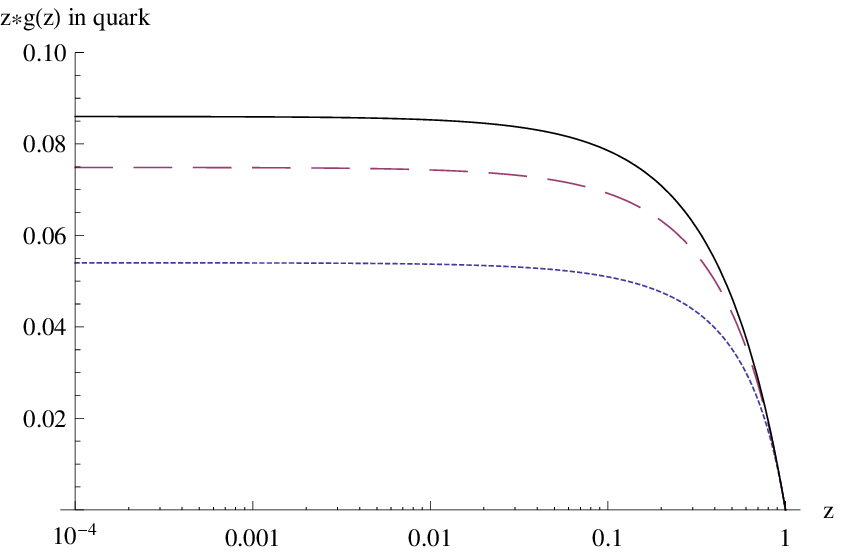}}\
%\hspace*{1.0cm}
 %\vskip -1cm
\end{minipage}
\begin{minipage}[c]{8cm}
\vskip -3cm \hspace*{8.0cm}
\centering{\includegraphics[width=5cm,height=3cm,angle=0]{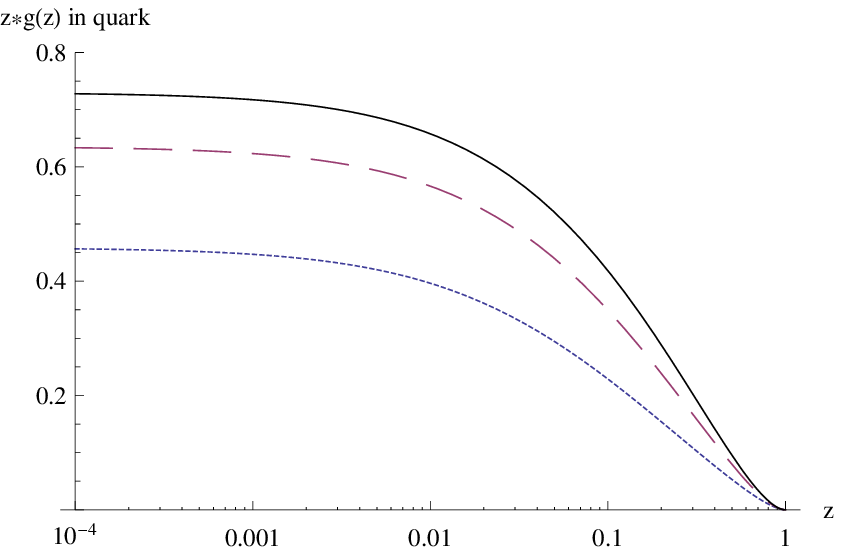}}\
%\hspace*{1.0cm}
\end{minipage}
\caption{ The $z$ dependency of the contributions to the the unpolarized gluon distribution
in the constituent quark from the pQCD (left panel),
from the non-perturbative interaction without pion (central panel), and from the non-perturbative
interaction with pion (right panel). The dotted line corresponds to $Q^2=2$ GeV$^2$,
the dashed line to $Q^2=5$ GeV$^2$, and the solid line to $Q^2=10$ GeV$^2$.}
\end{figure}
\end{widetext}

\begin{widetext}

\begin{figure}[h]
\begin{minipage}[c]{6cm}
%\vskip -0.5cm
\hspace*{-3.0cm}
\centering{\includegraphics[width=7cm,height=4cm,angle=0]{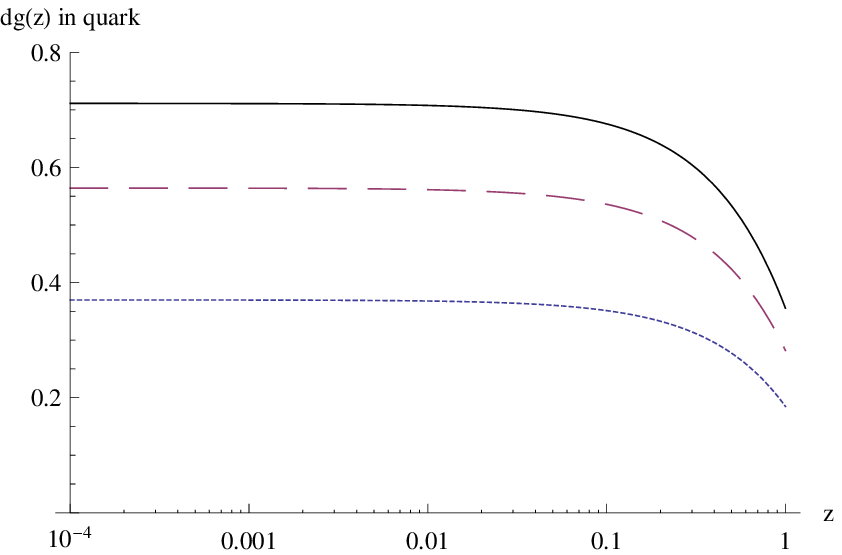}}\
\end{minipage}
\begin{minipage}[c]{6cm}
%\vskip -3cm
 \hspace*{1.0cm}
\centering{\includegraphics[width=7cm,height=4cm,angle=0]{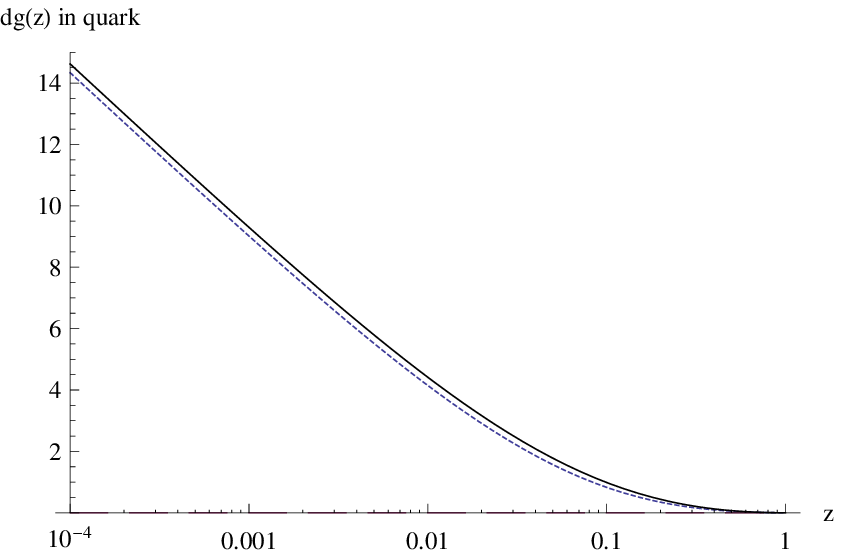}}\
%\hspace*{1.0cm}
\end{minipage}
\caption{ The $z$ dependency of the contributions to the polarized gluon distribution
in the constituent quark from the pQCD (left panel), and from the non-perturbative
interaction with pion (right panel). The notations are the same as in the Fig.2. In the right panel the result for
$Q^2=5 $ GeV$^2$ does not shown because it is practically identical to the $Q^2=10 $ GeV$^2$ case. }
\end{figure}
\end{widetext}
In the Fig.2 the results for the unpolarized gluon distribution
in constituent quark are presented as a function of the $z$ with the several $Q^2$.
One can see that the contribution from the pQCD to the unpolarized gluon distribution
has more hard $z$ dependency than the contributions from the nonperturbative interactions.
As the result, the pQCD gives the dominated contribution in the large $z$ region only.
The non-perturbative contribution coming from the diagram without pion, b) in Fig.1,
is rather small in comparison with the contributions from both the pQCD, a) in Fig.1,
and the non-perturbative interaction with pion, c) in Fig.1.
 One of the main reasons of the enhancement of the contributions coming
from the diagram with the pion is that it includes the additional integration over final particle momenta.
For the case of the one light  pion in the final state the available phase space is very large because of
$(M_q+m_\pi)^2<<E^2_{sph}$.
At small $z$ region all contributions exhibit Pomeron-like behavior, $zg(z,Q^2)\approx const$.
In the Fig.3 the results for the polarized gluon distributions in the constituent quark are shown.
Again one can see that the contribution from pQCD to the polarized gluon distribution dominates
in the region $z\rightarrow 1$ and the contribution from the non-perturbative interaction
with pion dominates in small $z$ region. The behaviors of two contributions in low $z$ region
are completely different. Namely, the contribution from pQCD shows the behavior as $\Delta g(z,Q^2)\rightarrow  const$,
on the other hand, the contribution from the non-perturbative interaction shows anomalous dependency
as $\Delta g(z,Q^2)\rightarrow \log(z)$. Additionally, their $Q^2$ dependencies are also very different.
The contribution from pQCD grows with $Q^2$ as $\log(Q^2\rho_c^2)$ but the contribution from
the non-perturbative interaction practically does not depend on the external $Q^2$ for  $Q^2>1/\rho_c^2=0.35$GeV~$^2$.
Therefore, we can treat the contribution from the nonperturbative interaction to the gluon distribution
as {\it an intrinsic polarized gluon} inside the constituent quark.

\section{Gluon  distributions in the nucleon}
We will apply the convolution model to obtain the gluon distributions in the nucleon
from the gluon distributions in the constituent quark.
Within this model the unpolarized and polarized gluon distributions in the nucleon are given by
\ba
g_N(x,Q^2)&=&\int_x^1\frac{dy}{y}q_V(y)g_q(\frac{x}{y},Q^2),\nonumber\\
\Delta g_N(x,Q^2)&=&\int_x^1\frac{dy}{y}\Delta q_V(y)\Delta g_q(\frac{x}{y},Q^2),
 \label{conv}
 \ea
where  $q_V(y)$ ($\Delta q_V(y)$) is unpolarized (polarized) distribution of
constituent quark in the nucleon and $g_q(z,Q^2)$ ($\Delta g_q(z,Q^2)$)
is the gluon distribution in the constituent quark obtained above.
For the unpolarized constituent quark distribution, we take
 \be
 q_V(y)=60y(1-y)^3.
\label{unpolvalence}
\ee
At the large $y$ this distribution is in accord with the quark counting
rule. Its behavior in small $y$ region and its normalization are fixed by the
requirements $\int_0^1dyq_V(y)=3$ and $\int_0^1dyyq_V(y)=1$.
It means that total momentum of nucleon at $Q^2\rightarrow 0 $ is carried by the three constituent quarks.
For the polarized constituent quark distribution the simple form is assumed
\be
 \Delta q_V(y)=2.4(1-y)^3.
\label{polvalence}
\ee
This form is also in  agreement with the  quark counting rule at $y\rightarrow 1$. The
normalization has been fixed from the hyperon weak decay data (see \cite{Aidala:2012mv}) as
\be
\int_0^1dy\Delta q_V(y)=\Delta u_V+\Delta d_V\approx 0.6.
\label{initialpol}
\ee

\begin{widetext}

\begin{figure}[h]
\begin{minipage}[c]{6cm}
%\vskip -0.5cm
\hspace*{-3.0cm}
\centering{\includegraphics[width=5cm,height=3cm,angle=0]{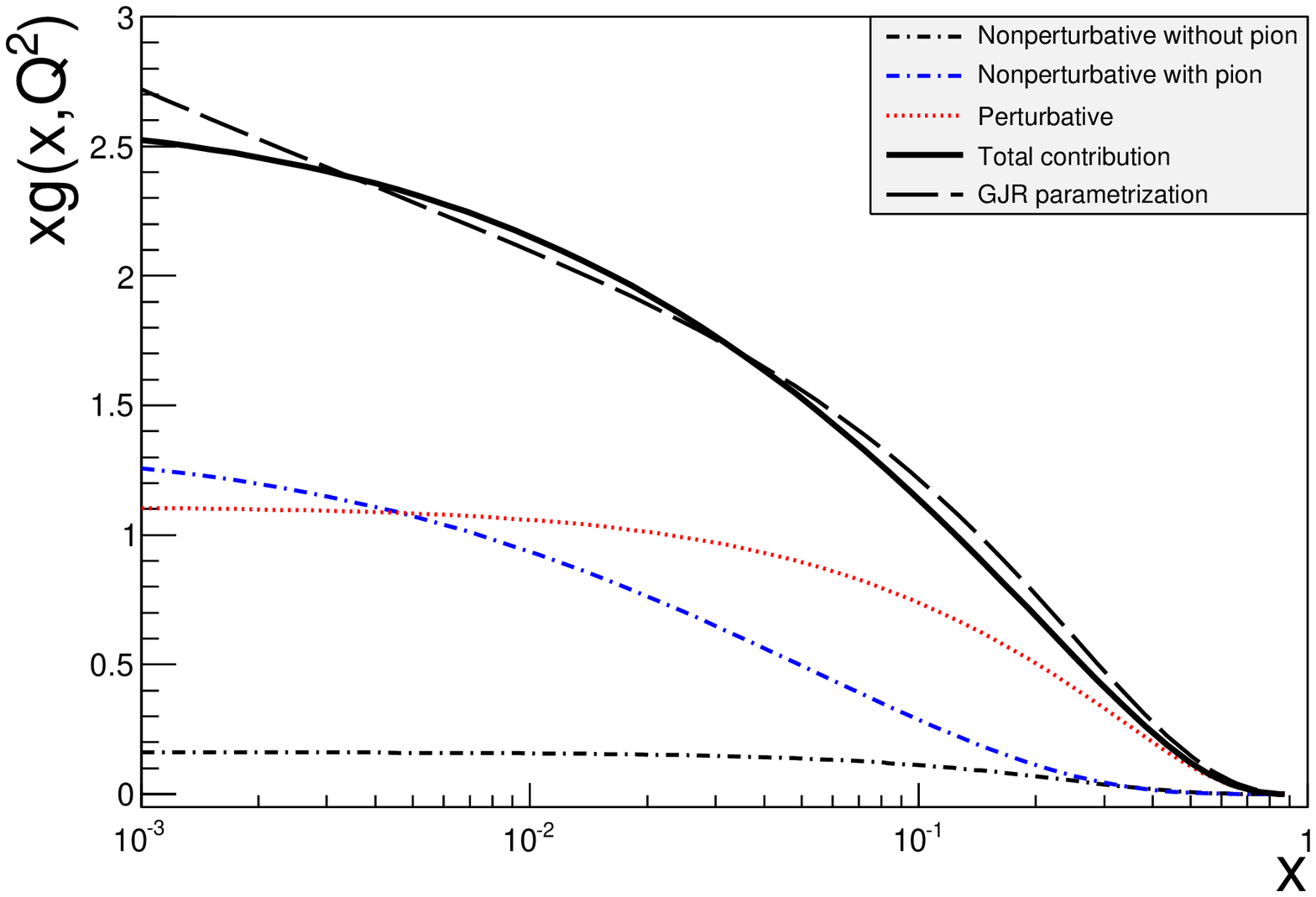}}\
\end{minipage}
\begin{minipage}[c]{6cm}
%\vskip -3cm
 \hspace*{2.0cm}
\centering{\includegraphics[width=5cm,height=3cm,angle=0]{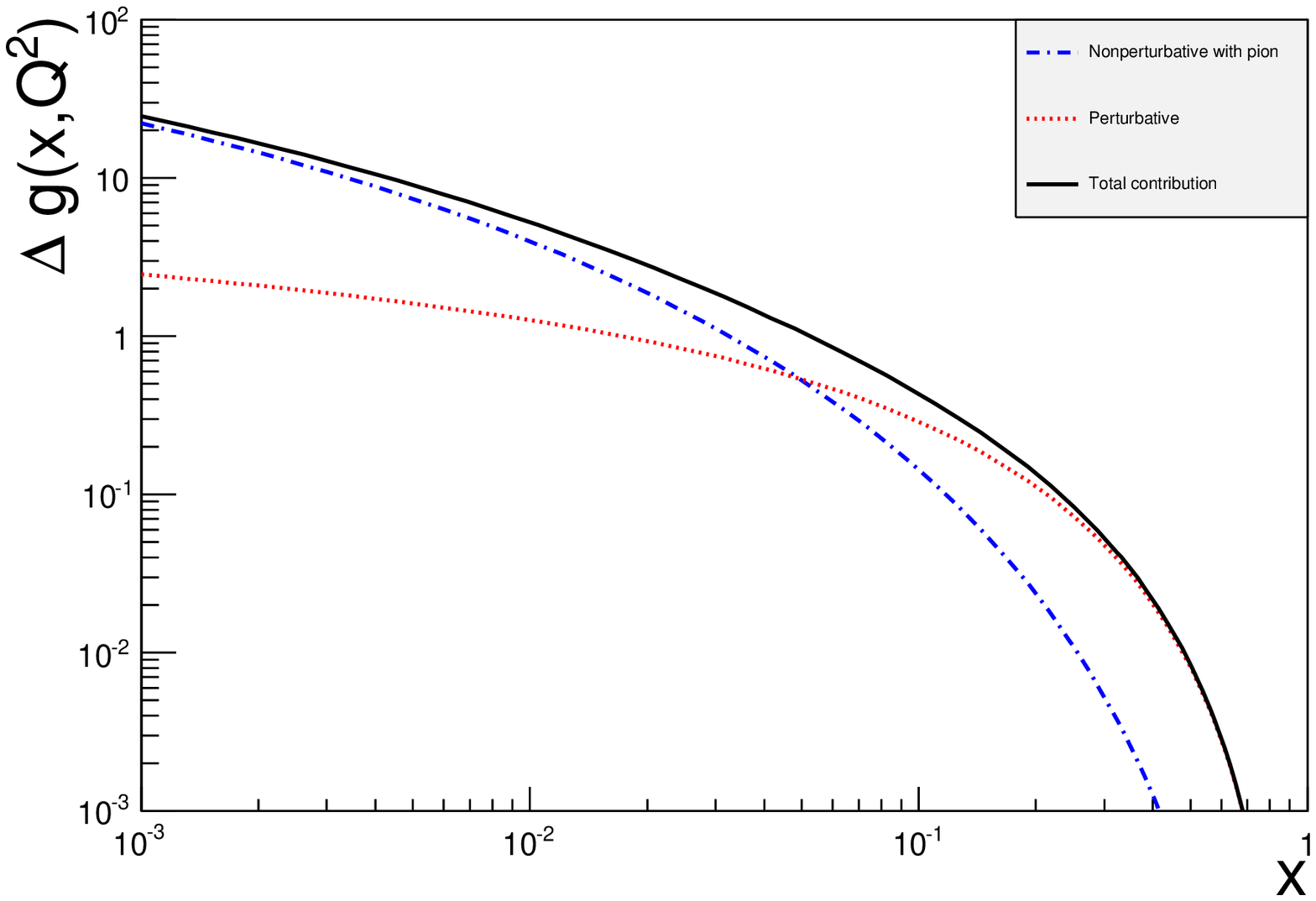}}\
%\hspace*{1.0cm}
\end{minipage}
\caption{ Unpolarized (left panel), and the  polarized
(right panel) gluon distributions in the nucleon at the scale $Q^2=2$ GeV$^2$. The dotted line in red corresponds to
the contribution from pQCD, the dotted-dashed in blue to the contribution from the non-perturbative interaction with pion,
dotted-dashed in black to the contribution from the non-perturbative interaction without pion,
and the solid line to the total contribution.}
\end{figure}
\end{widetext}
In the Fig.4, the result of the unpolarized and polarized gluon distributions in the nucleon
at the scale $Q_0^2=2$ GeV$^2$ is presented. This value of scale is often  used  as an input scale for the standard
pQCD evolution. Our result for the unpolarized gluon distribution in the left panel of Fig.4 is identical
to the GJR parametrization $ xg(x)=1.37x^{-0.1}(1-x)^{3.33} $~\cite{Gluck:2007ck}.
It is well known that the behavior of non-polarized gluon distribution in low $x$ region is determined
by the Pomeron exchange. This exchange plays very important role in the phenomenology of high energy reactions.
Our results in Figs.2,4 indicate the existence of two different Pomerons. One is  the "hard" pQCD Pomeron
and another is the "soft" non-perturbative Pomeron with quite different dependency on $x$ and $Q^2$.
The existence of the two types of the Pomerons can explain in the natural way simultaneously both the DIS data at large $Q^2$
and the high energy cross sections with small momentum transfer~(see \cite{Landshoff:2009wt} and references therein).
In the right panel of Fig.4 the result for the polarized gluon distribution is shown.
The $Q^2$ dependency of the part of the nucleon momentum carried by the gluons $G(Q^2)=\int_0^1dxxg(x,Q^2)$
and their polarization $\Delta G(Q^2)=\int_0^1dx \Delta g(x,Q^2)$ are presented in Fig.5.
It is clear that the non-perturbative interaction without pion gives a small contribution to
unpolarized gluon distribution. Furthermore, as it was mentioned above, such contribution is zero
for the polarized gluon case. So, for $Q^2>1/\rho_c^2$, the main contributions to both the unpolarized
and the polarized gluon distributions are come from pQCD and from the non-perturbative interaction with the pion.

\begin{widetext}

\begin{figure}[h]
\begin{minipage}[c]{6cm}
%\vskip -0.5cm
\hspace*{-3.0cm}
\centering{\includegraphics[width=5cm,height=3cm,angle=0]{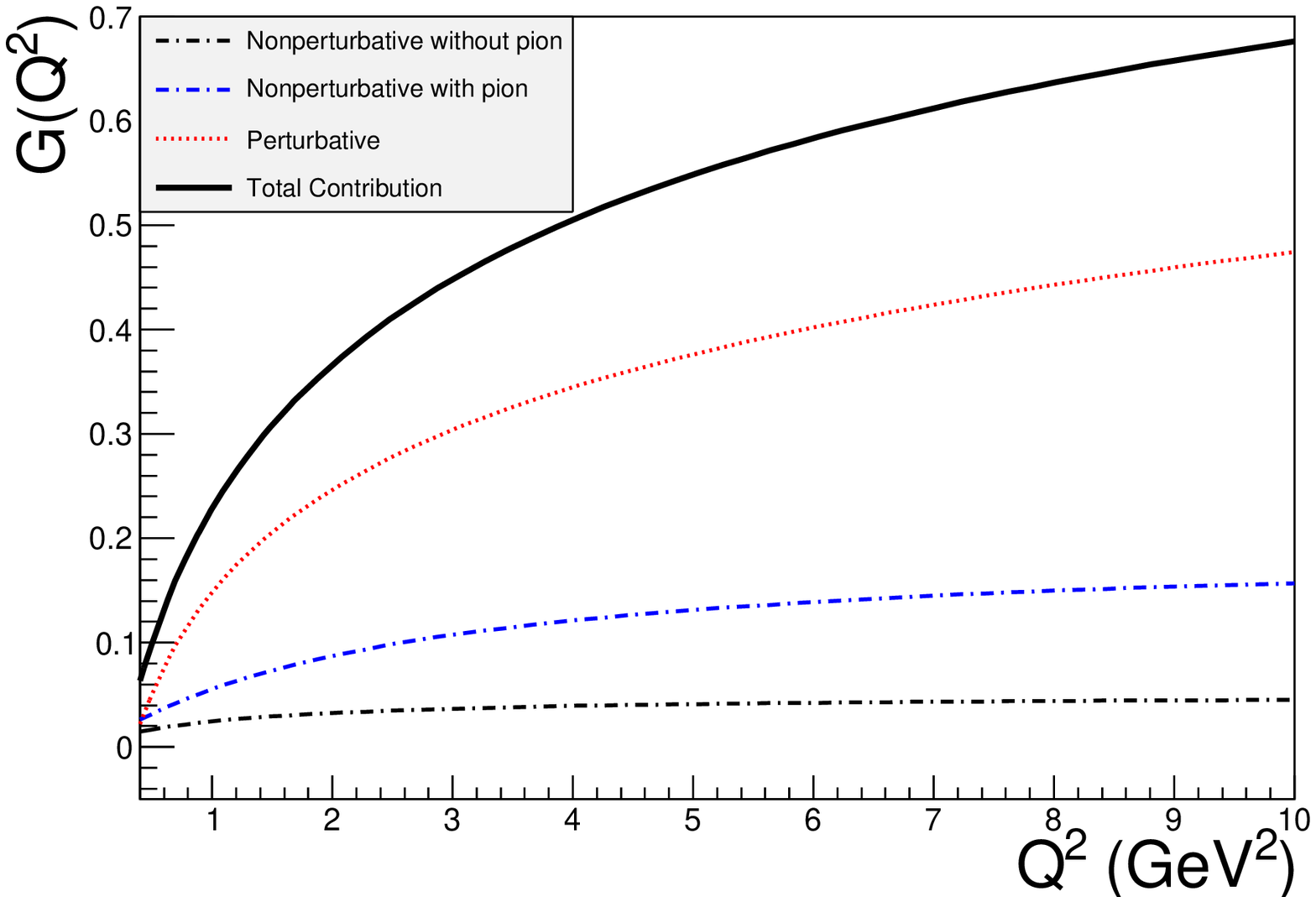}}\
\end{minipage}
\begin{minipage}[c]{6cm}
%\vskip -3cm
 \hspace*{2.0cm}
\centering{\includegraphics[width=5cm,height=3cm,angle=0]{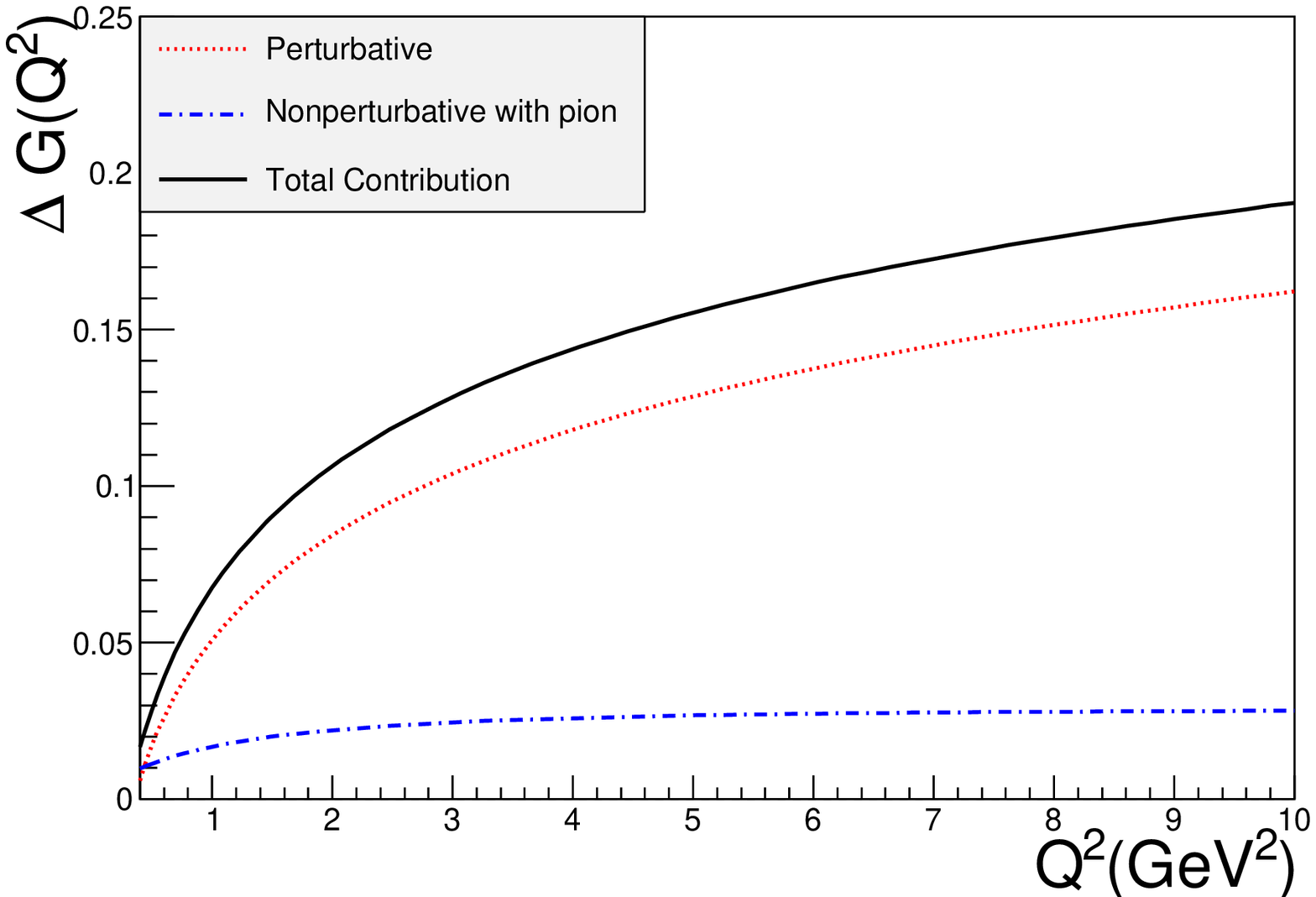}}\
%\hspace*{1.0cm}
\end{minipage}
\caption{ The part of the nucleon momentum carried by gluons (left panel), and the contribution of the gluons
to nucleon spin (right panel) as the function of $Q^2$.
The notations are the same as in the Fig.4 }
\end{figure}
\end{widetext}

The famous proton spin problem is one of the longstanding puzzle in the  QCD \cite{Aidala:2012mv}.
The gluon polarization might be very important part in the following decomposition of the proton spin
by Jaffe and Manohar \cite{Jaffe:1989jz}
\be
\frac{1}{2}=\frac{1}{2}\Delta\Sigma+\Delta G+L_q+ L_g,
\label{spin}
\ee
where the first term is quark contribution, $\Delta G=\int_0^1dx\Delta g(x)$ is the gluon polarization in nucleon
and the last two terms are the contributions from the orbital motions of the quarks and the gluons.
The main problem is how to explain very small value of the proton spin carried by quark.
At the present, the typical value is $\Delta\Sigma \approx 0.25$ \cite{Aidala:2012mv},
which is far away from $\Delta\Sigma=1$ given by the non-relativistic quark model.
The relativistic motion of the quarks in the confinement region results in sizable decreasing of total helicity
of quarks. For example, within the bag model one obtains $\Delta\Sigma=0.65$.
We should point out that this value is in agreement with the weak hyperon decay data, Eq.\ref{initialpol},
but it is not enough to explain the small value coming from deep inelastic scattering (DIS)
at large $Q^2\geq 1$ GeV$^2$. Just after appearance of the EMC data on the small part of the spin of
the proton carried by the quarks, the axial anomaly effect in DIS was considered as the primary effect
to solve this problem \cite{Altarelli:1988nr}.
For the three light quark flavors, it gives the following reduction of the quark helicity in the DIS
\be
\Delta\Sigma_{DIS}=\Delta\Sigma-\frac{3\alpha_s}{2\pi}\Delta G
\label{DIS}
 \ee
It is evident that one needs to have a huge
positive gluon polarization, $\Delta G\approx 3\div 4$, in the proton to explain the small value of the  $\Delta\Sigma_{DIS}$.
The modern experimental data from the inclusive hadron productions and the jet productions exclude such a
large gluon polarization in the accessible intervals of $x$ and $Q^2$ \cite{deFlorian:2014yva,Nocera:2014gqa}.
Our model also exclude the large polarization of gluons (see Fig.5, right panel)
in these intervals. For example, at $Q^2=10$ GeV$^2$ the $\int_0^1dx \Delta g(x)=0.19$ in our model.
This value is in agreement with recent fit of available data on spin asymmetry from the jet productions
and the inclusive hadron productions  in  \cite{deFlorian:2014yva,Nocera:2014gqa}.
Therefore, the axial anomaly effect, suggested in \cite{Altarelli:1988nr}, cannot explain the proton spin problem.
We should stress that the helicity of the initial quark is flipped in the vertices b) and c) in Fig.1.
As the result, such vertices should lead to the {\it screening} of the quark helicity. It is evident that at the $Q^2\rightarrow 0$
such screening is vanished as shown in Fig.5 and the total spin of the proton  is carried by its constituent quarks.
Due to the total angular momentum conservation, for $Q^2\neq 0$, the flip of the quark helicity by the non-perturbative interactions
should be compensated partially by the orbital momenta of the partons and pion {\it inside} the constituent quark.
In addition to such compensation, there should be a compensation coming from the positive gluon polarization
inside the constituent quark. Similar instanton induced mechanism to solve the
proton spin crisis was discussed many years ago on qualitatively in the papers
\cite{Dorokhov:1993fc,Dorokhov:1993ym}. The careful calculation of such effect based on the Lagrangian
Eq.\ref{Lag} is in the progress.\\
Let us discuss some possible  uncertainties in our calculation.
In the calculation we  neglected the high order  terms in the pion field.  Their contribution to the
cross section is expected to be suppressed in a large $N_c$ limit
 by factor $1/N_c$, because $F_\pi\sim \sqrt{N_c}$. Additionally,
multi-pion contribution
should  be suppressed  by the restriction in the  phase space which is coming from the
the value of the height of the instanton barrier $E\approx 2.7$ GeV. For example, the two pion production
one can consider as a  massive sigma meson production which should be strongly suppressed by the available
phase space.
In any case, it is evident that multi-pion contribution will only enhance the pion contribution to the gluon PDFs
and cannot give the influence to our final conclusion about dominated role of the pion in the generation of the gluons
in the hadrons.
We would like to mention that our approach, which is based on the effective Lagrangian Eq.\ref{Lag2},
in some sense is dual to the approaches which are using the current quark propagator
in instanton background  and/or  multi-gluon
instanton induced vertices to calculate some high energy cross sections
\cite{Kharzeev:2000ef,Moch:1996bs,Schrempp:2002kd,Shuryak:2003rb}. In our model all soft gluon effects
are integrated out and their effects are included inside the  effective interaction.
It is happened that our  approach is much  simpler and  more effective in the calculation of such important
ingredients for the high energy cross sections as the gluon distributions which
 have never been obtained by other methods  based on the instanton model for QCD vacuum. Furthermore, it can be also applied
 to calculate  the spin-dependent high energy cross-sections in
the spirit of the papers \cite{Kochelev:2013zoa,Qian:2015wyq,Ostrovsky:2004pd}.

\section{Conclusion}
We calculate the gluon distributions in the constituent quark and in the nucleon.
Our approach are based on the anomalous quark-gluon and quark-gluon-pion interactions induced
by non-trivial topological structure of the QCD vacuum.
It is shown that the quark-gluon-pion anomalous interaction gives a very large contribution to  both
the unpolarized and polarized gluon distributions. It means that pion field plays a fundamental
role to produce  both gluon distributions in hadrons.
The possibility of the matching of the constituent quark model for the nucleon with
its partonic picture is shown. The phenomenological arguments in favor of such a non-perturbative
gluon structure of the constituent quark were recently given in the papers \cite{Kopeliovich:2007pq,Kopeliovich:2006bm}.
We also pointed out that the famous proton spin crisis might be explained by the
flipping of the helicity of the quark induced by non-perturbative anomalous quark-gluon and quark-gluon-
pion interactions.

\section{Acknowledgments}
We are grateful to Igor Cherednikov and Aleksander Dorokhov for useful discussions.
 This work was partially supported by the National Natural
  Science Foundation of China (Grant No. 11575254 and 11175215), and by the Chinese Academy of Sciences visiting
professorship for senior international scientists (Grant No.
2013T2J0011). This research was also supported in part by the
Basic Science Research Program through the National Research
Foundation of Korea(NRF) funded by the Ministry of
Education(2013R1A1A2009695)(HJL).


\begin{thebibliography}{99}
\bibitem{Aidala:2012mv}
  C.~A.~Aidala, S.~D.~Bass, D.~Hasch and G.~K.~Mallot,
  %``The Spin Structure of the Nucleon,''
  Rev.\ Mod.\ Phys.\  {\bf 85}, 655 (2013).

 \bibitem{shuryak} T. Sch\"afer and E.V. Shuryak,
Rev. Mod. Phys. {\bf 70} (1998) 1323.

\bibitem{diakonov}
  D.~Diakonov,
  %``Instantons at work,''
  Prog.\ Part.\ Nucl.\ Phys.\  {\bf 51} (2003) 173.



   \bibitem{kochelev3}
  N.~Kochelev and N.~Korchagin,
  %``Anomalous Quark Chromomagnetic Moment and Single-Spin Asymmetries,''
    Phys.\ Lett.\ B {\bf 729}, 117 (2014).




  \bibitem{kochelev1}
   N.~I.~Kochelev,
   Phys.\ Lett.\  {\bf B426} (1998) 149.

   \bibitem{kochelev4}
  N.~Kochelev,
  %``Role of anomalous chromomagnetic interaction in pomeron and odderon structures and in gluon distribution,''
   Phys.\ Part.\ Nucl.\ Lett.\  {\bf 7}, 326 (2010).


  \bibitem{Balla:1997hf}
  J.~Balla, M.~V.~Polyakov and C.~Weiss,
  %``Nucleon matrix elements of higher twist operators from the instanton vacuum,''
  Nucl.\ Phys.\ B {\bf 510}, 327 (1998).


  \bibitem{Kochelev:2015pha}
  N.~Kochelev, H.~J.~Lee, B.~Zhang and P.~Zhang,
  %``Anomalous pion production induced by nontrivial topological structure of QCD vacuum,''
  Phys.\ Rev.\ D {\bf 92}, no. 3, 034025 (2015).

  \bibitem{Kochelev:2015jba}
  N.~Kochelev, H.~J.~Lee, Y.~Oh, B.~Zhang and P.~Zhang,
  %``Non-perturbative collisional energy loss of heavy quarks in quark-gluon plasma,''
  arXiv:1510.00472 [hep-ph].







\bibitem{Faccioli:2001ug}
  P.~Faccioli and E.~V.~Shuryak,
  %``Systematic study of the single instanton approximation in QCD,''
  Phys.\ Rev.\ D {\bf 64}, 114020 (2001).

\bibitem{Altarelli:1977zs}
  G.~Altarelli and G.~Parisi,
  %``Asymptotic Freedom in Parton Language,''
  Nucl.\ Phys.\ B {\bf 126}, 298 (1977).


\bibitem{Kochelev:1997qq}
  N.~I.~Kochelev,
  %``Instanton contribution to polarized and unpolarized gluon distribution in nucleon,''
  hep-ph/9707418.

  \bibitem{Gluck:2007ck}
  M.~Gluck, P.~Jimenez-Delgado and E.~Reya,
  %``Dynamical parton distributions of the nucleon and very small-x physics,''
  Eur.\ Phys.\ J.\ C {\bf 53}, 355 (2008).

\bibitem{Landshoff:2009wt}
  P.~V.~Landshoff,
  %``Fundamental problems with hadronic and leptonic interactions,''
  Acta Phys.\ Polon.\ B {\bf 40}, 1967 (2009)


\bibitem{Jaffe:1989jz}
  R.~L.~Jaffe and A.~Manohar,
  %``The G(1) Problem: Fact and Fantasy on the Spin of the Proton,''
  Nucl.\ Phys.\ B {\bf 337}, 509 (1990).

  \bibitem{Altarelli:1988nr}
 A.V. Efremov and O.V. Teryaev, Dubna preprint E2-88-287 (1988);
  G.~Altarelli and G.~G.~Ross,
  %``The Anomalous Gluon Contribution to Polarized Leptoproduction,''
  Phys.\ Lett.\ B {\bf 212}, 391 (1988).


\bibitem{deFlorian:2014yva}
  D.~de Florian, R.~Sassot, M.~Stratmann and W.~Vogelsang,
  %``Evidence for polarization of gluons in the proton,''
  Phys.\ Rev.\ Lett.\  {\bf 113}, no. 1, 012001 (2014).

  \bibitem{Nocera:2014gqa}
  E.~R.~Nocera {\it et al.} [NNPDF Collaboration],
  %``A first unbiased global determination of polarized PDFs and their uncertainties,''
  Nucl.\ Phys.\ B {\bf 887}, 276 (2014).


  \bibitem{Dorokhov:1993fc}
  A.~E.~Dorokhov and N.~I.~Kochelev,
  %``Instanton induced asymmetric quark configurations in the nucleon and parton sum rules,''
  Phys.\ Lett.\ B {\bf 304}, 167 (1993).

  \bibitem{Dorokhov:1993ym}
  A.~E.~Dorokhov, N.~I.~Kochelev and Y.~A.~Zubov,
  %``Proton spin within nonperturbative QCD,''
  Int.\ J.\ Mod.\ Phys.\ A {\bf 8}, 603 (1993).


  \bibitem{Kopeliovich:2007pq}
  B.~Z.~Kopeliovich, I.~K.~Potashnikova, B.~Povh and I.~Schmidt,
  %``Evidences for two scales in hadrons,''
  Phys.\ Rev.\ D {\bf 76}, 094020 (2007).


  \bibitem{Kopeliovich:2006bm}
  B.~Z.~Kopeliovich, B.~Povh and I.~Schmidt,
  %``Glue drops inside hadrons,''
  Nucl.\ Phys.\ A {\bf 782}, 24 (2007).


  \bibitem{Kharzeev:2000ef}
  D.~E.~Kharzeev, Y.~V.~Kovchegov and E.~Levin,
  %``QCD instantons and the soft pomeron,''
  Nucl.\ Phys.\ A {\bf 690}, 621 (2001).


  \bibitem{Moch:1996bs}
   S.~Moch, A.~Ringwald and F.~Schrempp,
  %``Instantons in deep inelastic scattering: The Simplest process,''
  Nucl.\ Phys.\ B {\bf 507}, 134 (1997)

\bibitem{Schrempp:2002kd}
  F.~Schrempp and A.~Utermann,
  %``QCD instantons and high-energy diffractive scattering,''
  Phys.\ Lett.\ B {\bf 543}, 197 (2002).

 \bibitem{Shuryak:2003rb}
  E.~V.~Shuryak and I.~Zahed,
Phys.\ Rev.\ D {\bf 69}, 014011 (2004).

\bibitem{Kochelev:2013zoa}
  N.~Kochelev and N.~Korchagin,
  %``Anomalous Quark Chromomagnetic Moment and Single-Spin Asymmetries,''
  Phys.\ Lett.\ B {\bf 729}, 117 (2014)


\bibitem{Qian:2015wyq}
  Y.~Qian and I.~Zahed,
  %``Spin Physics through QCD Instantons,''
  arXiv:1512.08172 [hep-ph].


\bibitem{Ostrovsky:2004pd}
  D.~Ostrovsky and E.~Shuryak,
  %``Instanton-induced azimuthal spin asymmetry in deep inelastic scattering,''
  Phys.\ Rev.\ D {\bf 71}, 014037 (2005).




\end{thebibliography}
\end{document}